# Magnetoelectricity at *room temperature* in $Bi_{0.9-x}Tb_xLa_{0.1}FeO_3$ system

V. R. Palkar[1*], Darshan C. Kundaliya,[1,2] S.K. Malik[1] and S. Bhattacharya[1]

[1]Department of Condensed Matter Physics & Materials Science,
Tata Institute of Fundamental Research, Mumbai 400 005, India
[2]Center for Superconductivity Research, Department of Physics, University of Maryland,
College Park, Maryland-20742, U.S.A.

## Abstract

Magnetoelectric compounds with the general formula, $Bi_{0.9-x}R_xLa_{0.1}FeO_3$ (R =Gd, Tb, Dy, etc.), have been synthesized. These show the coexistence of ferroelectricity and magnetism, possess high dielectric constant and exhibit magnetoelectric coupling at *room temperature*. Such materials may be of great significance in basic as well as applied research.

PACS No: 75.50D, 75.80, 77.84

*corresponding author
e-mail: palkar@tifr.res.in

In some insulating materials, an external magnetic field can induce electric polarization and an external electric field can induce changes in magnetization. This phenomenon is known as the magnetoelectric (ME) effect and materials exhibiting this effect are called magnetoelectrics or signetto-magnets. Thus magnetoelectric materials show the coexistence of ferroelectric and magnetic ordering and, therefore, could be used either as ferroelectrics or as magnetic materials in devices since both the properties exist simultaneously. Ability to couple to either the electric or the magnetic polarization allows an additional degree of freedom in device design. However, the ME effect is restricted to insulators and is not so large in magnitude. Further, most of the materials reported so far show ME effect at temperatures much below room temperatures[1.2]. Recently, Wang *et al*[3] reported an enhancement of polarization and related properties in heteroepitaxially constrained thin films of $BiFeO_3$. However, the observed enhancement in magnetization is thickness dependent (up to 75nm) and is not a bulk property in factual sense. Hence the search for single compounds exhibiting magnetoelectric effect at and above room temperature continues.

During the course of our research on ferroelectric and magnetic materials, we have been successful in synthesizing new compounds showing ME effect at *room temperature*. The new compounds are based on bismuth ferrite ($BiFeO_3$) - a compound which is known to possess a ferroelectric ordering (transition temperature, $T_C$~840°C) and an antiferromagnetic ordering (Néel temperature, $T_N$~340°C) at room temperature. We have synthesized and studied the properties of $Bi_{0.9-x}R_xLa_{0.1}FeO_3$ compounds, where R could be Gd, Tb, Dy, etc., and x varies between 0.05-0.3 (a small amount of La is added to stabilize[4] the perovskite phase of $BiFeO_3$). In this communication, we report the synthesis, and crystal structure of $Bi_{0.9-x}Tb_xLa_{0.1}FeO_3$ compounds, and ferroelectric and ferromagnetic properties of one representative compound, namely, $Bi_{0.825}Tb_{0.075}La_{0.1}FeO_3$. We also demonstrate the presence of magnetoelectric coupling, at room temperature, in this compound.

The $Bi_{0.9-x}Tb_xLa_{0.1}FeO_3$ powder samples ($0 \leq x \leq 0.3$) were synthesized by using a novel wet chemical route developed in our laboratory. All the cations were co-precipitated as hydroxides by using NaOH as precipitating agent. Hydroxy complex thus obtained was calcined at 550ºC to get reacted materials. More preparation details are described elsewhere[5]. The reacted samples have been characterized by various techniques. X-ray powder diffraction (XRD) was used for phase identification. Differential Thermal Analysis (DTA) was carried out to determine magnetic and ferroelectric transition temperatures ($T_M$ and $T_C$, respectively). Capacitance of the samples was measured by using LCZ meter (HP Model 4277A) at a frequency of 10 kHz and field amplitude of 20mVrms. Ferroelectric hysteresis behavior was studied by using **a** RT66 Loop Tracer (Radiant Technology) in Sawyer-Tower mode. Top and bottom surfaces of the sintered pellet were coated with silver and used as electrodes. **A** SQUID magnetometer [MPMS, Quantum Design] was used for measuring the field dependence of the magnetization of the samples.

The compound $BiFeO_3$ crystallizes in the rhombohedrally distorted perovskite structure. From XRD studies, we find that the Tb substituted $Bi_{0.9-x}Tb_xLa_{0.1}FeO_3$ compounds are also single-phase materials crystallizing in the same structure as the parent $BiFeO_3$ compound. A typical x-ray pattern for the sample with x=0.075 is shown in Fig. 1. However, Tb and La substitution has not affected the crystalline structure of the parent compound and this is important for the ferroelectric properties of the compounds. However, Tb and La substitution leads to a decrease in the unit cell volume from 186(Å)$^3$ for x=0 to 183.1(Å)$^3$ for x=0.075. The decrease in unit cell volume with increasing Tb substitution is expected since the ionic radius of $Tb^{+3}$ (0.923Å) is smaller than that of $Bi^{+3}$ (0.96Å). Figure 2 shows the results of DTA measurements on $BiFeO_3$ and $Bi_{0.825}Tb_{0.075}La_{0.1}FeO_3$ samples in (a) 200ºC to 400ºC temperature interval and (b) in 780ºC to 850ºC temperature interval. For the parent compound, the peak obtained in DTA at ~ 340ºC (Fig. 2(a)) is known to be due to magnetic ordering in this compound. The same is shifted down to 245ºC in the Tb substituted compound. Incidentally, a dielectric anomaly has been also observed near the magnetic transition temperature (Fig. 3). This type of dielectric anomaly in magnetoelectrically ordered systems is predicted by the Landau-Devonshire theory of phase transition as an influence of vanishing magnetic order on the electric order[6]. The peak at ~806°C in DTA (Fig. 2(b)) is attributed to ferroelectric transition and occurs at a slightly lower temperature than that in parent $BiFeO_3$[7,8]. The decrease in ferroelectric transition temperatures ($T_C$) of $BiFeO_3$ on Tb substitution may be ascribed to the decrease in unit cell volume caused by Tb.

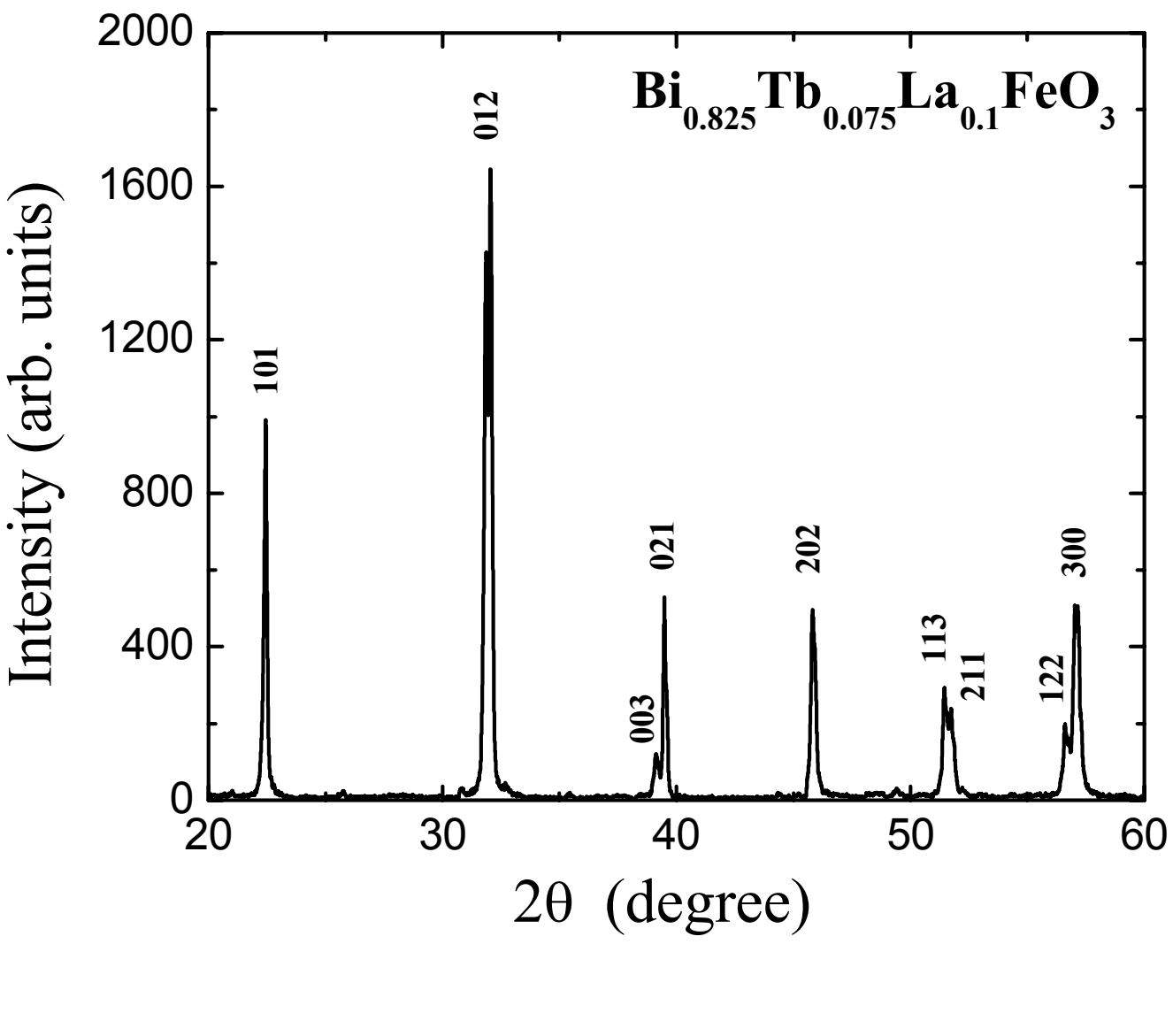

Fig. 1. XRD pattern of $Bi_{0.825}Tb_{0.075}La_{0.1}FeO_3$ sample indicating the absence of impurity phases

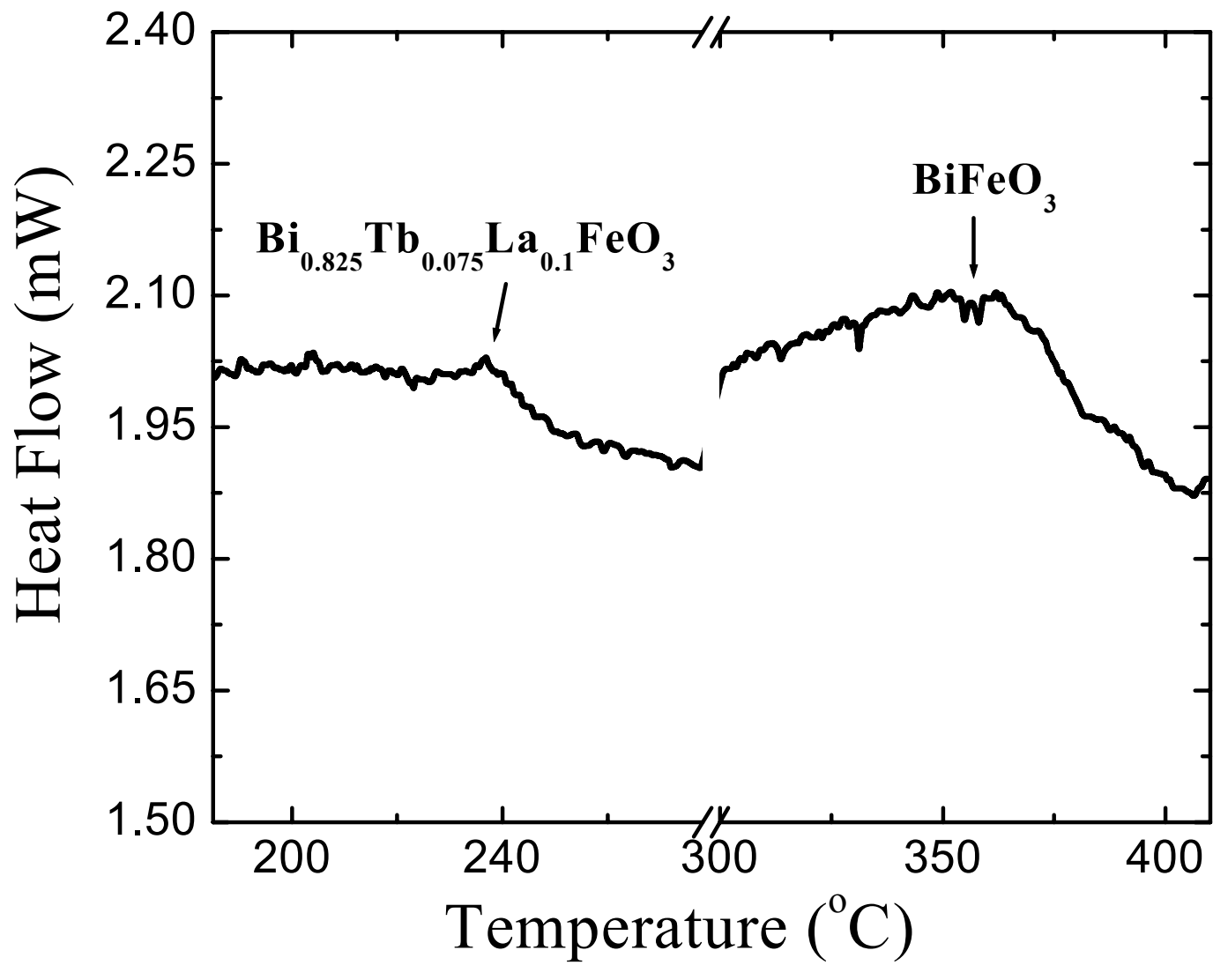

Fig. 2(a) DTA curve obtained for $Bi_{0.825}Tb_{0.075}La_{0.1}FeO_3$ and $BiFeO_3$ samples showing the change in magnetic ordering ($T_M$) temperature

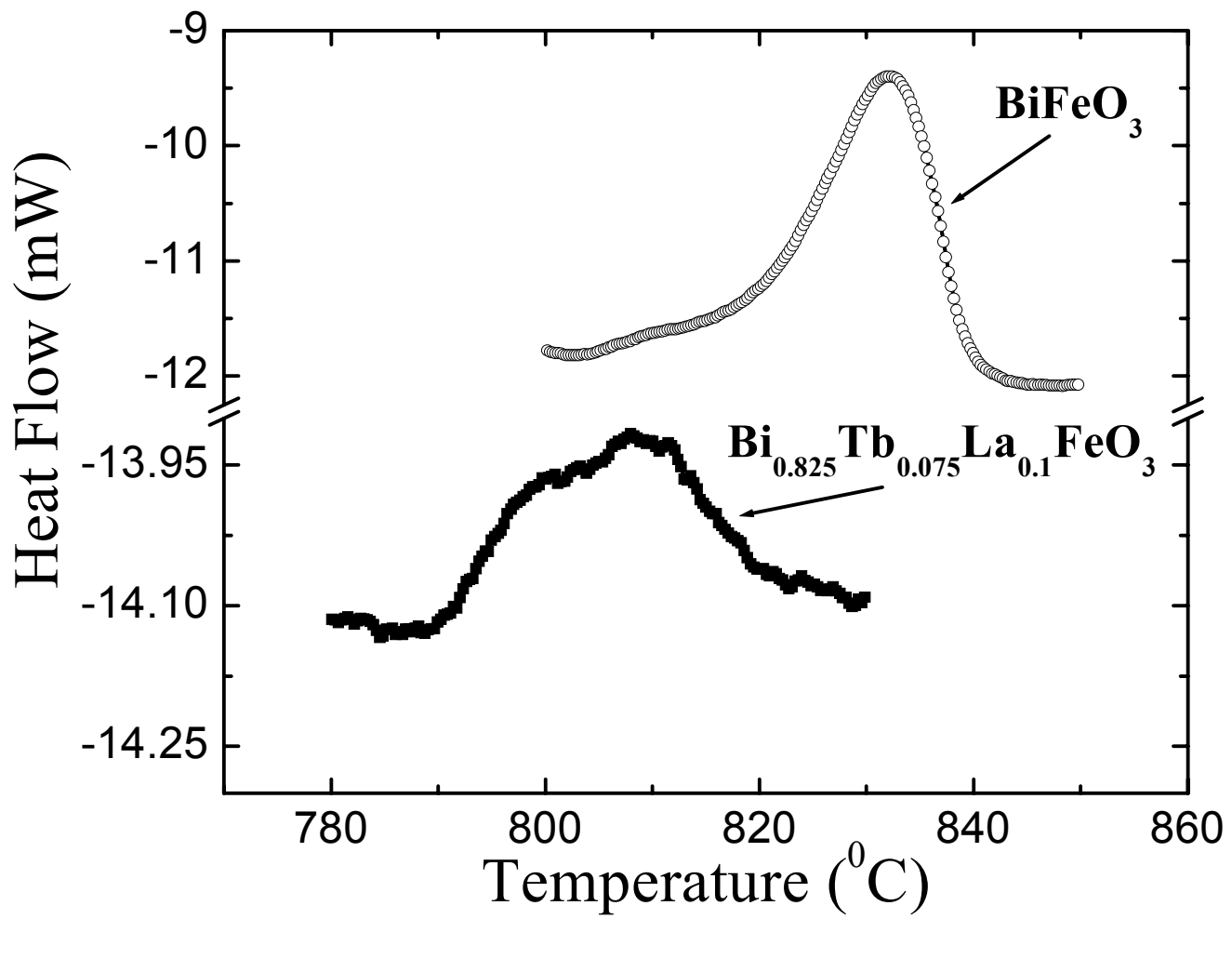

Fig. 2(b) DTA curve obtained for $Bi_{0.825}Tb_{0.075}La_{0.1}FeO_3$ and $BiFeO_3$ samples showing the change in ferroelectric transition ($T_C$) temperature.

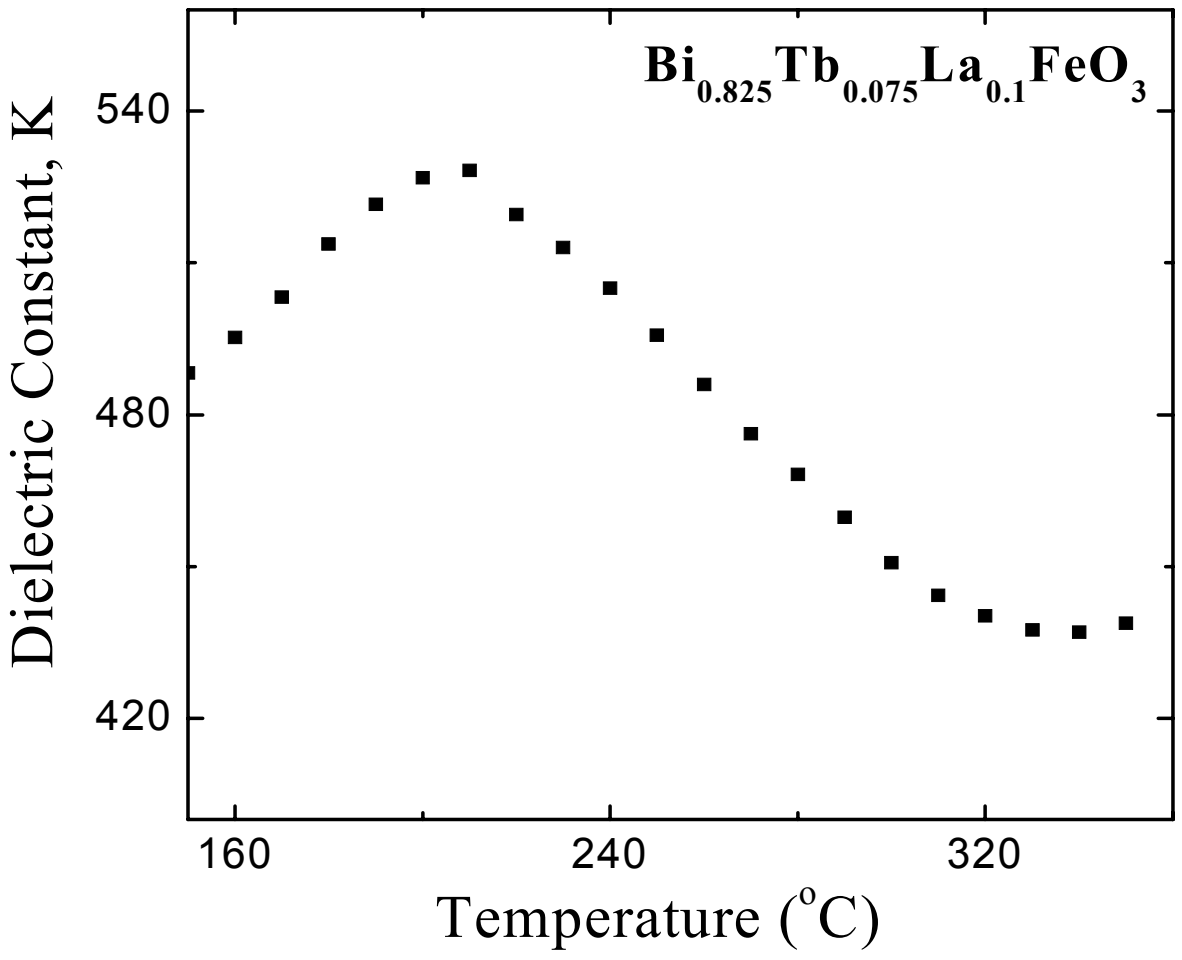

Fig. 3. Dielectric response with temperature measured for $Bi_{0.825}Tb_{0.075}La_{0.1}FeO_3$ sample.

The polarization ($P_s$) versus applied electric field (E) hysteresis loop for as-sintered $Bi_{0.825}Tb_{0.075}La_{0.1}FeO_3$ pellet is shown in Fig. 4. Saturated hysteresis loop seen in this figure reveals the ferroelectric behavior of the as-sintered pellet. Figure 5 shows the variation of magnetization (M) with applied field (H) at room temperature for $BiFeO_3$ as well as for $Bi_{0.825}Tb_{0.075}La_{0.1}FeO_3$ samples. The magnetization for pure $BiFeO_3$ is small and varies linearly with field as expected for an antiferromagnetic material. In contrast, a saturation in the M-H curve is obtained at 300K in $Bi_{0.825}Tb_{0.075}La_{0.1}FeO_3$ which indicates that the sample is ferro- or ferrimagnetically ordered even at room temperature, consistent with the magnetic ordering temperature of 245°C observed in DTA measurements. Thus ferroelectric- and magnetic -orders coexist in this compound at room temperature. The ferromagnetic moment at 300K is estimated to be about $4\mu_B$ per Tb ion which may increase at lower temperatures and tend towards $Tb^{3+}$ free ion value.

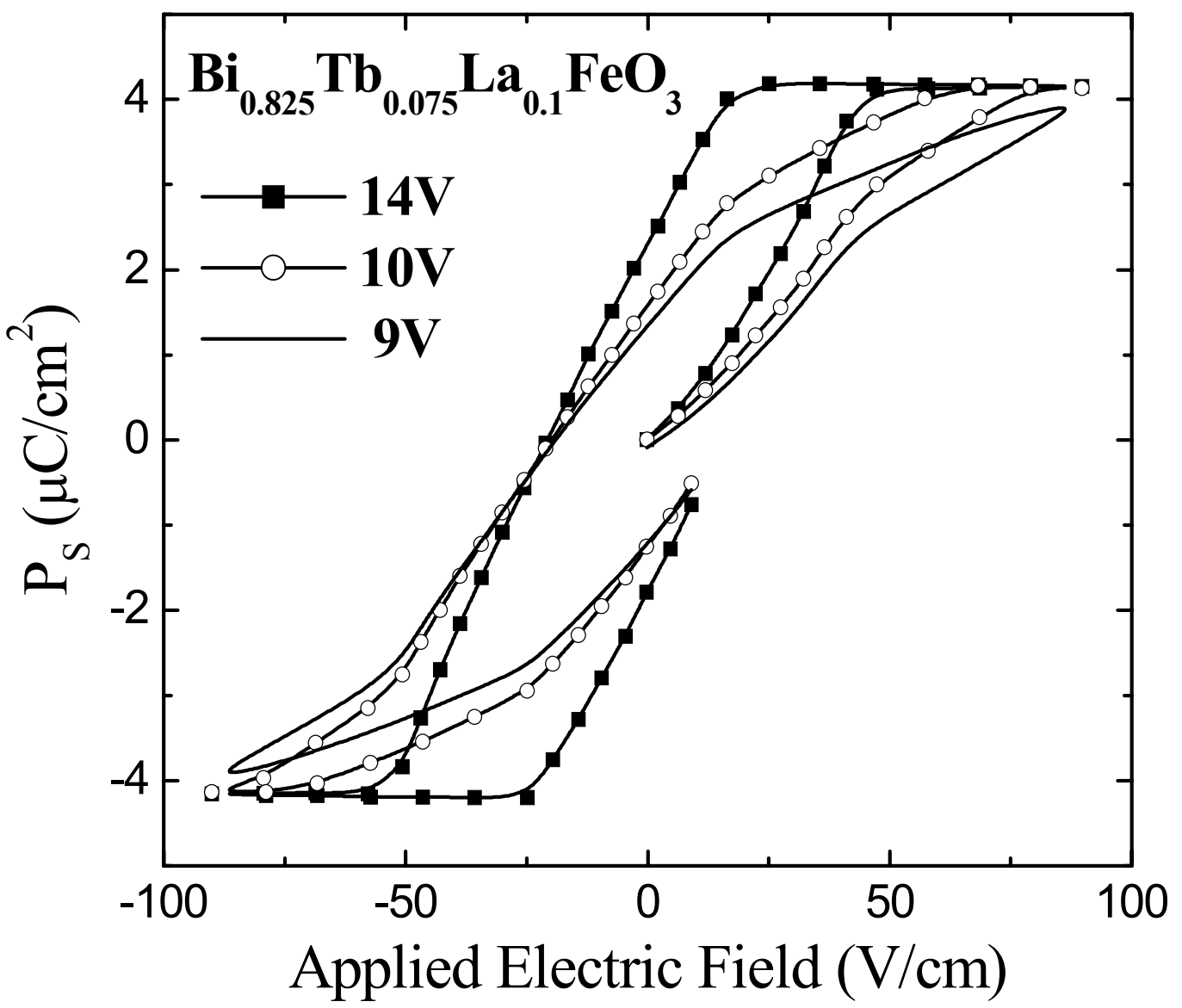

Fig. 4. Saturated ferroelectric hysteresis loop obtained on $Bi_{0.825}Tb_{0.075}La_{0.1}FeO_3$ pellet sample. Resistivity of the sample is ~$10^8$ Ω.cm

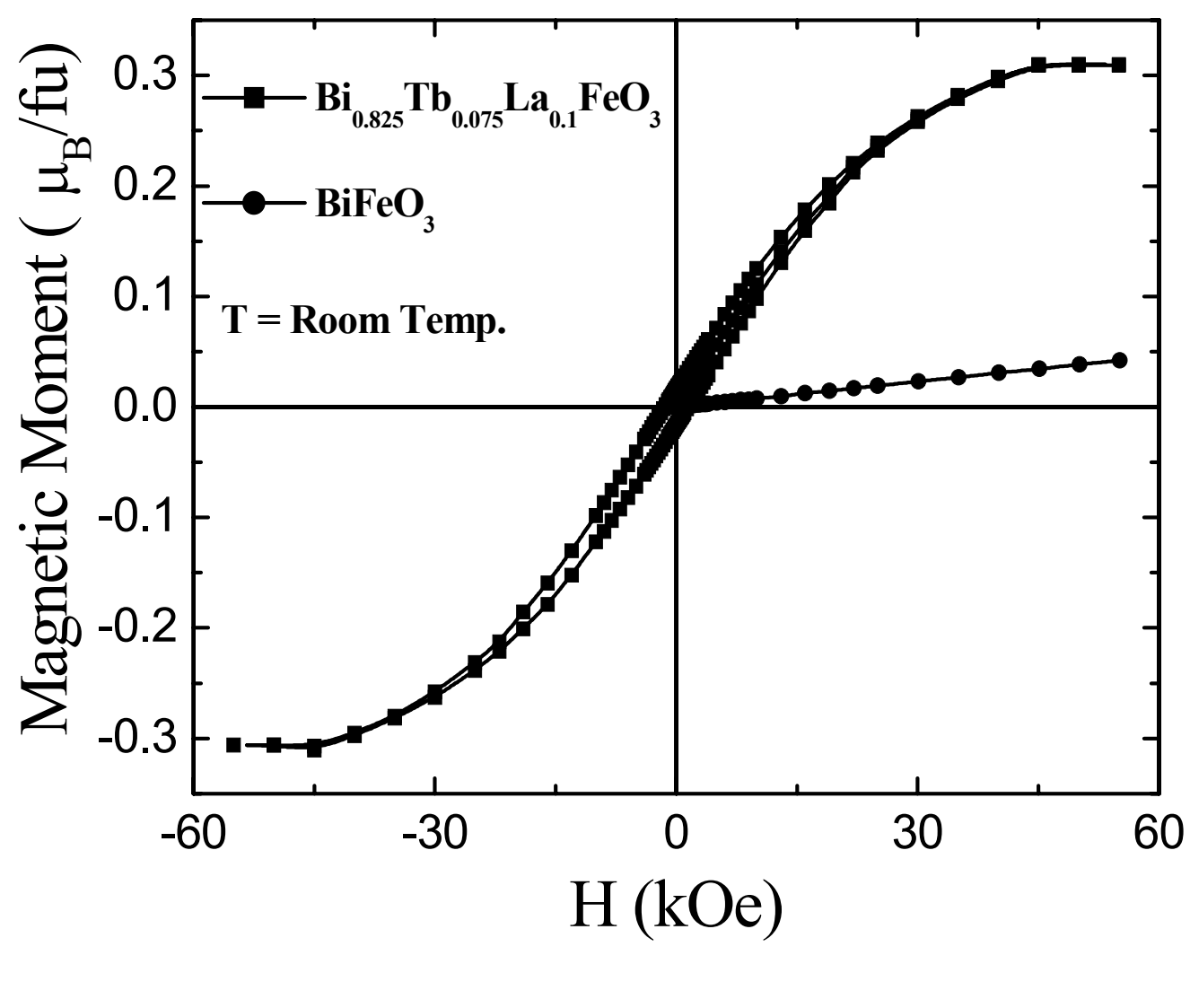

Fig. 5. M-H curve obtained at room temperature for $Bi_{0.825}Tb_{0.075}La_{0.1}FeO_3$ and $BiFeO_3$ samples.

In order to demonstrate the coupling between electric and magnetic polarizations in $Bi_{0.825}Tb_{0.075}La_{0.1}FeO_3$, we have carried out two experiments, namely, (a) measurement of electric polarization after poling the sample in a magnetic field, and (b) *in situ* changes in capacitance with application of a magnetic field. Figure 6 shows the change in saturation polarization ($P_s$) values for the sample after poling at different magnetic fields. A continuous increase in $P_s$ is seen with increase in poling field before $P_s$ eventually saturates. This is indicative of the coupling between the two polarizations. An increase in dielectric constant with increase in magnetic field is observed (Figure 7) and this is also a signature of the ME conversion occurring in the sample. The magnetoelectric coupling observed at room temperature and the observed reduction in lattice parameters indirectly confirm that Tb is not remaining isolated in the matrix of the compound but enters the lattice. Although the theory of magnetoelectric media[9] is based on symmetry arguments, which are not yet available for our novel ferromagnetic-ferroelectric compound, the essence of magnetoelectric coupling may be envisaged as follows. When a magnetic field is applied to a magnetoelectric material, the material is strained. This strain induces a stress on the piezoelectric (all ferroelectrics are piezoelectrics), which generates the electric field. This field could orient the ferroelectric domains leading to an increase in polarization value. Such magnetoelectric coupling and the large dielectic constant observed in the present system could be useful in device applications.

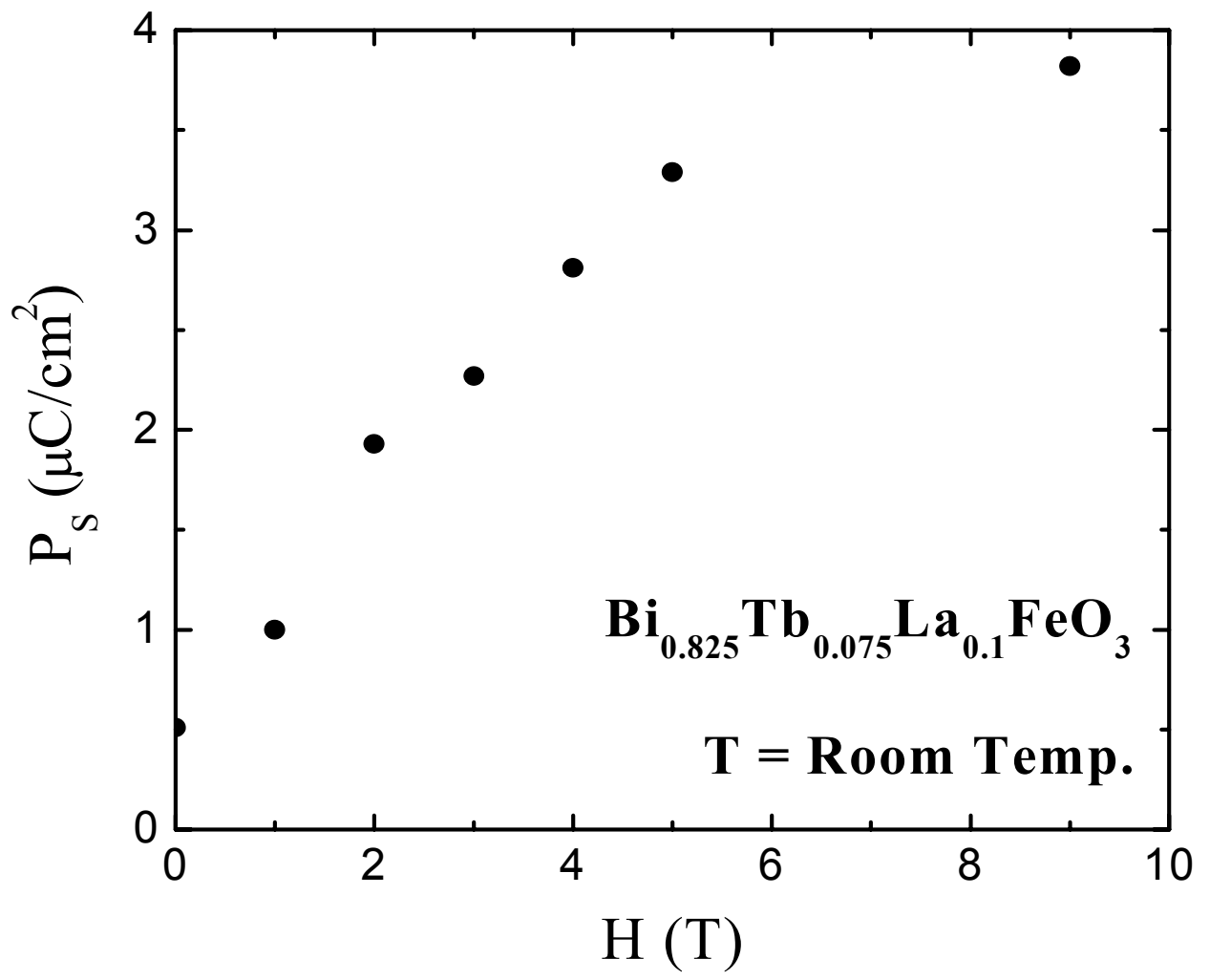

Fig. 6. Saturation polarization ($P_S$) vs. magnetic field used for poling $Bi_{0.825}Tb_{0.075}La_{0.1}FeO_3$ pellet sample.

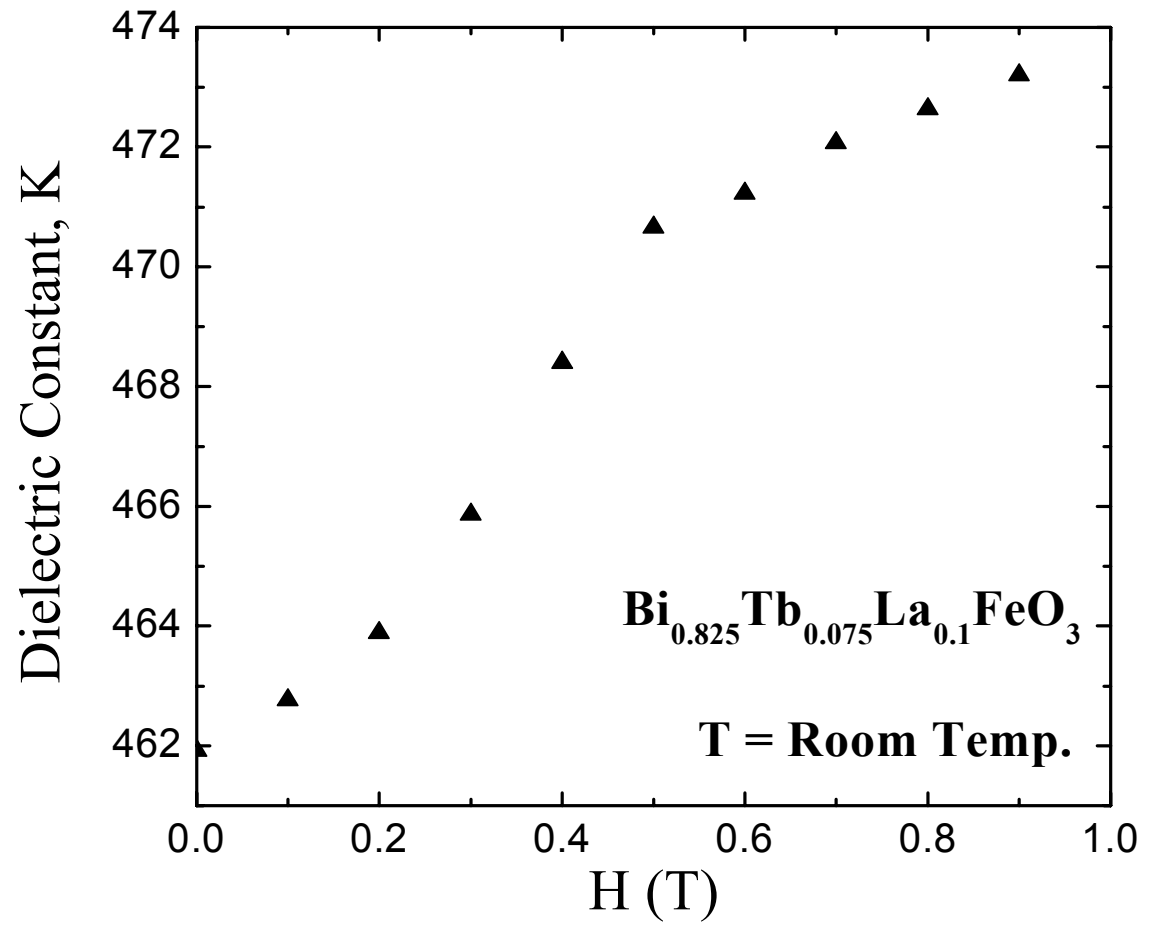

Fig. 7. Dielectric constant, K vs. magnetic field (H) curve for $Bi_{0.825}Tb_{0.075}La_{0.1}FeO_3$ pellet sample.

In summary, we have been successful in synthesizing a new system exhibiting magnetoelectric coupling and large dielectric constant at room temperature. The material characteristics may be highly useful in numerous devices. We have also been able to grow thin films of the same material with similar properties and the result will be published elsewhere.

The authors wish to thank A.V. Gurjar, J. John, Ms. B.A. Chalke and D. Buddhikot for experimental help.